\documentclass[a4paper]{jpconf}
\usepackage{graphicx}
\usepackage{iopams}
\usepackage{amsmath}

\begin{document}
\title{\large Presented at \\Physics Beyond Relativity 2019 conference \\ Prague, Czech Republic, October 20, 2019
\\ (invited presentation)  \\ \vspace{2cm} \huge Reference Frame Transformations and Quantization}

\author{Akira Kanda$^1$, Renata Wong$^2$ and Mihai Prunescu$^3$}

\address{$^1$ Omega Mathematical Institute/University of Toronto, Toronto, Ontario, Canada} 
\address{$^2$ Department of Computer Science and Technology, Nanjing University, Nanjing, China} 
\address{$^3$ University of Bucharest, Bucharest, Romania}

\ead{kanda@cs.toronto.edu, renata.wong@protonmail.com, mihai.prunescu@gmail.com}
\begin{abstract}
It has been said that Maxwell's theory of electromagnetic field is relativistic as Einstein showed that these axioms of Maxwell
are all Lorentz invariant. We investigate some issues regarding these results. 

``Mozart, You are a god, and do not even know it." ... (Alexander
Pushkin)

``To punish my contempt of authority, I became an authority." ...
(Albert Einstein)
\end{abstract}

\section{Transformations and dynamics}
One of the biggest and most fundamental questions in relativity
theory is what does it mean that something is ``relativistic". Newton assumed the absolute reference frame and defined relative motion
as the difference of two absolute motion vectors to be observed from the outside
of the absolute frame. Galileo's idea was to reject the absolute frame and
associate a reference frame to each body and let each body observe the other
bodies' motion inside its own reference frame. This concept came with
the restriction of ``inertial reference frames", by which are meant the reference
frames that move with constant speed relative to each other inside each
other. This condition was added when it was discovered that accelerating
reference frames do not share the same laws of physics. For example in the
classical dynamics, by considering the mutually accelerating frames, we
violate the law of action-reaction. The condition of sharing the same laws of
physics was taken as fundamental for the theory of relativity and is called the
``principle of relativity".

To represent the motion of inertial reference frames, Galilean
relativity theory introduced a naturally associated ``spacetime" coordinate
transformation called Galilean transformation:
\begin{eqnarray*}
t^{\prime } &=&t \\
x^{\prime } &=&x-vt
\end{eqnarray*}
where $v$ is the relative speed between the two inertial
frames involved. With this, it was possible to restate the principle
of relativity as ``The laws of physics must be invariant under the Galilean
transformation."

Knowing that the adoption of inertial reference frames
violates the third law (the law of action-reaction) of Newton's dynamics, it seems a natural
question to ask whether the second law and the law of gravity are invariant under the
Galilean transformation. Curiously, this question was never considered.
However, as we will present below, the answer is ``yes" for Galilean
transformation.
\begin{equation*}
F=m\frac{d^{2}x}{dt^{2}}\quad \implies \quad F=m\frac{d^{2}(x-vt)}{dt^{2}}=m%
\frac{d^{2}x}{dt^{2}}.
\end{equation*}
\begin{equation*}
F=\frac{GmM}{(x_{m}^{{}}-x_{M})^{2}}\quad \implies \quad F=\frac{GmM}{%
((x_{m}-vt)^{{}}-(x_{M}^{{}}-vt))^{2}}=\frac{GmM}{(x_{m}-x_{M}^{{}})^{2}}.
\end{equation*}
Hence, in the case of Galilean transformation, the damage inflicted by relativization
is limited to the loss of the third law.

\textbf{Remark (1)}
\textit{By identical argument we can show that the Coulomb force law is
invariant under the Galilean transformation.}

To the Galilean relativity theory Einstein added an extra axiom of the constancy of the speed of light, which says that the speed of light is
constant $c$ in any inertial reference frames, in other words $c+v=c$.\cite{Ein} The result is what we now call the special theory of relativity. The simplest and most effective refutation of this claim came from Anderton \cite{And} of Natural Philosophy Alliance. Anderton incisively pointed out that ``if $c+v=c$ is true then ${\normalsize c}$ is not a speed." We present the following argument to back up Anderton's argument. 
Assume that $v$ is the absolute speed of Michelson-Morley
apparatus in the absolute space. Even if $c+v$ is the absolute
speed of the light moving towards the mirror, the effect of this cancels out
because the mirror is also moving with speed $v$ in the absolute
frame. On the same token, though the reflected light moves with speed $c-v$ towards the emitter of light, as the emitter is moving with speed 
$v$, the effect of $v$ cancels out. So, we will never
detect this $v$. This came from an incorrect interpretation of
Michelson-Morley experiment.

Thus, Einstein derived the Lorentz transformation between two inertial reference frames. Before Einstein, Lorentz derived a coordinate transformation only between the absolute aether frame and an observer's frame moving in the absolute frame, assuming that the length of a matter moving in the absolute frame contracts (also referred to as ``Lorentz-FitzGerald contraction").\cite{Lor} Einstein generalized this result of Lorentz to the setting of arbitrary inertial reference frames without involving the absolute frame. It goes as follows:
\begin{eqnarray*}
x\prime  &=&\frac{(x-vt)}{\sqrt{1-(v/c)^{2}}} \\
t^{\prime } &=&\frac{1}{\sqrt{1-(v/c)^{2}}}\left( t-\frac{vx}{c^{2}}\right) .
\end{eqnarray*}
With this, Einstein rewrote the principle of relativity to read that ``laws of
physics must be invariant under the choice of inertial reference frames", which later was taken as the invariance under the Lorentz transformations.
Before this the principle of relativity meant that the laws of physics must
be invariant under the choice of inertial reference frames. (Though not
brought up explicitly, this implied that the laws of physics
must be invariant under the Galilean transformations.)

Being the extension of the Galilean
relativity theory, this theory violates the third law of dynamics. Interestingly, also the second law and the law of gravity fail to be
invariant under the Lorentz transformation, as we have
\begin{equation*}
F=m\frac{d^{2}x}{dt^{2}}\quad \implies \quad F=m\frac{d^{2}}{dt^{2}}\frac{%
(x-vt)}{\sqrt{1-(v/c)^{2}}}\neq m\frac{d^{2}x}{dt^{2}}.
\end{equation*}
\begin{equation*}
F=\frac{GmM}{(x_{m}^{{}}-x_{M})^{2}}\Longrightarrow F=\frac{GmM}{(\frac{%
(x_{m}-vt)}{\sqrt{1-(v/c)^{2}}}^{{}}-\frac{(x_{M}-vt)}{\sqrt{1-(v/c)^{2}}}%
)^{2}}=\frac{GmM}{(\frac{(x_{m}-x_{M})}{\sqrt{1-(v/c)^{2}}})^{2}}\neq \frac{%
GmM}{(x_{m}^{{}}-x_{M})^{2}}.
\end{equation*}
This means that under Einsteinian relativisation all axioms of
Newtonian dynamics except the first axiom of the Law of Inertia fail. The
damage inflicted by relativisation is much bigger when we take Einsteinian
relativisation over Galilean relativisation. 

\textbf{Remark (2)}
\textit{Again, by identical argument we can show that Lorentz transformation
fails to conserve the Coulomb's laws.}

\section{Transformations and electromagnetism}
One of the major reasons for the introduction of the Lorentz
transformation was that Lorentz discovered that the electromagnetic wave
equation of Maxwell is not invariant under the Galilean transformation. Lorentz
discovered that a better coordinate transformation by himself, called
Lorentz transformation, preserves Maxwell's electromagnetic wave equation.\cite{Lor} Einstein generalized this result by showing that
under the Lorentz transformation all basic equations of Maxwell's
electromagnetic field theory are invariant.\cite{Ein} This became the
vindication for the claim that Maxwell's electromagnetic field theory is
relativistic in Einsteinian sense.

Though wave equations are not conserved under the Galilean
transformations, wave functions are transformed into wave functions through
Galilean transformations. So, it is not quite clear if we needed Lorentz
transformations to begin with. If the issue of relativity is just the
changing of reference frames, then Galilean transformation is the most
natural transformation representing the choice of inertial reference frames.
The dominating argument for supporting the Lorentz transformation is twofold:

\begin{enumerate}
\item Lorentz transformation maps electromagnetic wave equations
into electromagnetic wave equations while Galilean transformations fail to
do so. The right response to this is that wave mechanics and particle
mechanics are entirely different theories covering entirely different issues
of physics. Galilean transformation came from the consideration of
relativising particle kinematics. It is therefore totally expected that this
transformation does not deal with physical waves, which exist upon continuum
wave mediums. In wave mechanics, it is not particles that move in the
direction of the wave. It is the localized vibration energy of the medium
that moves. For this reason, wave mechanics has no naturally associated
momentum, which is the product of speed and mass. The hypothetical momentum
of waves came from the relativistic theory of waves of de
Broglie.\cite{DeB}

\item Lorentz transformation transforms Maxwell's electromagnetic
equations into equivalent equations. We will discuss later the issue of whether Galilean transformations do the same.
\end{enumerate}

Maxwell reduced the entire electromagnetic theory to the following
field plus current equations:
\begin{equation*}
\mathbf{\nabla \times E}=-\frac{1}{c}\frac{\partial \mathbf{H}}{\partial t}%
,\quad \mathbf{\nabla \cdot E}=4\pi \rho ,\quad \mathbf{\nabla \cdot H}%
=0,\quad \mathbf{J}=\rho \mathbf{v,\quad \nabla \times H}=\frac{4\pi }{c}%
\mathbf{J}+\frac{1}{c}\frac{\partial }{\partial t}\mathbf{E}
\end{equation*}
where $c=\frac{1}{\sqrt{\epsilon _{0}\mu _{0}}}.$ Here $\rho $ is electric charge density and $\mathbf{J}$ is the
conducting current created by the charge density moving
with speed $\mathbf{v}$. Maxwell obtained the last equation, called
the ``Generalized Amp\`{e}re's Law", under the assumption that the charges in $\mathbf{J}$ move with constant speed $\mathbf{v}$. This restriction was removed later.

This introduction of $\mathbf{J}$ into the field theory is
problematic. According to the definition of electric force field, charges
placed in a field will not affect the field, meaning that the charges
placed will not affect other charges that create the electric field.

Ontologically, it is hard to understand $\mathbf{J}$ at
this basic level. How is it possible that mutually repelling electrons can
form a coherent bundle? A possible answer says that this is possible inside a conductor. But there is a
vicious circularity here. Conductors are objects to be studied in material
science where we have to use the theory of electromagnetism. This problem somehow
resembles the problem of tension rope in classical dynamics. The rope is
supposed to be a mass-less entity that moves under acceleration. Then it must
be the case that all such ropes will move with infinite speed. How does it
connect two bodies? 

Here is yet another consequence of the logical inconsistency
discussed above of Maxwell's electromagnetic field theory. Maxwell obtained
the electromagnetic wave equation
\begin{equation*}
\mathbf{\nabla }^{2}\mathbf{E=}\frac{1}{c^{2}}\frac{\partial ^{2}\mathbf{E}}{%
\partial ^{2}t}\text{\qquad \textrm{and}\qquad }\mathbf{\nabla }^{2}\mathbf{%
H=}\frac{1}{c^{2}}\frac{\partial ^{2}\mathbf{H}}{\partial ^{2}t}\mathbf{.}
\end{equation*}
under the condition that $\mathbf{J}=0$ (meaning that there is no
conducting current). From this wave equation, he calculated the speed of
electromagnetic wave in vacuum to be $c=1/\sqrt{\varepsilon _{0}\mu _{0}}.$ 
Interestingly, Maxwell also showed that electromagnetic waves are created by
accelerating charges. But do accelerating charges not constitute a current?
According to the mathematical definition of a current, even a singular moving charge is a current. In fact, radio
engineers postulate that a circular closed circuit conducting
electrons inside produces electromagnetic wave and one cycle of the
electron corresponds to the cycle of the produced electromagnetic wave. This is because
electrons in the circuit are under centripetal acceleration.

Lorentz found out that the Galilean
transformation of the electromagnetic wave equations does not lead to wave equations. This lead him
to try his own Lorentz transformation instead and he showed that the Lorentz
transformation maps electromagnetic wave equation to electromagnetic wave
equation. 

\textbf{Remark (3)}
\textit{This claim by Lorentz is to be refuted later in \textbf{Section 6} however.}

The above formed a base for the claim that Galilean transformation is
invalid and should be replaced by Lorentz transformation, despite all other
problems that Lorentz transformations create, some of which we discussed in the
previous section.

\section{Transformation of waves v.s. transformation of wave
equations}
The argument that Galilean transformation is
invalid and should be replaced by Lorentz transformation should be reconsidered due to the
following legitimate argument: The wave equation in one dimension without
sources for speed $v$ is
\begin{equation*}
\mathbf{\nabla }^{2}\phi \text{ }\mathbf{=}\frac{1}{k}\frac{\partial
^{2}\phi }{\partial ^{2}t}.
\end{equation*}
It is well known that when we Galilean transform this equation the
result is not a wave equation anymore.

The general solution of this wave equation is
\begin{equation*}
\phi (x,t)=\psi _{+}(x-kt)+\psi _{-}(x+kt).
\end{equation*}
The first term is a wave propagating with speed $+k$, and
the second one with $-k$. The Galilean transformation transforms
this general solution to
\begin{equation*}
\phi (x,t)=\psi _{+}(x-(k+v)t)+\psi _{-}(x+(k-v)t).
\end{equation*}
This is a superposition of two waves and so it is a wave. This wave
has a different speed from the original wave. But is it not expected due to
the Galilean transformation?

We can support this argument using Fourier expansion too. A
sinusoidal wave is transformed into a sinusoidal wave. So, a Fourier expansion
which describes a wave also transforms into a Fourier series which
represents a wave.

However, when we Lorentz transform a wave function, we do not obtain a
wave function in general. For example it is known well that a sinusoidal wave
will be transformed into a sinusoidal wave by Lorentz transformation if
the wave amplitude is the solution of a wave equation which is invariant
under the Lorentz transformation.

For this reason, relativistic theory of
waves assumes that the wave equations are invariant under the Lorentz
transformation. Under this assumption, we have the invariance under Lorentz transformation of the phase of a plane wave
\begin{equation*}
\mathbf{k}^{\prime }\cdot \mathbf{r}^{\prime }-\omega ^{\prime }t^{\prime }=%
\mathbf{k}\cdot \mathbf{r}-\omega t
\end{equation*}
where $\mathbf{k}$ is the wave vector, $\mathbf{r}$ is a position vector and $\omega $ is the frequency.
With this invariance, we obtain the following relativistic wave transformation
equations:
\begin{eqnarray*}
k_{x}^{\prime } &=&\frac{1}{\sqrt{1-(v/c)^{2}}}\left( k_{x}-v\frac{\omega }{%
c^{2}}\right)  \\
k_{y}^{\prime } &=&k_{y} \\
k_{z}^{\prime } &=&k_{z} \\
\omega ^{\prime } &=&\frac{1}{\sqrt{1-(v/c)^{2}}}\left( \omega
-vk_{x}\right) .
\end{eqnarray*}
So it is clear that the theory of Lorentz transformation of waves
is not so universal. It applies only to
the waves that are invariant under the transformation. And it has been claimed 
that electromagnetic waves are examples of such waves. Clearly it has the same kind of deficiency as the Galilean transformation of waves, if not worse.

\textbf{Remark (4)}
\textit{It is clear that the phase $\mathbf{k}\cdot \mathbf{r}-\omega t$ is Galilean invariant for any wave.}

\textbf{Remark (5)}
\textit{We will show later in \textbf{Section [6]} that, contrary to what has
been claimed by Lorentz and Einstein, Lorentz transformations will not
conserve any of wave equations, electromagnetic or non-electromagnetic.}

De Broglie used this transformation
of ``relativistic" waves and Einstein's relativistic theory of photons to
create a relativistic wave-particle duality. Observing an analogy
between the above discussed relativistic wave transformation and the
relativistic transformation of energy and momentum:
\begin{eqnarray*}
p_{x}^{\prime } &=&\frac{1}{\sqrt{1-(v/c)^{2}}}\left( p_{x}-v\frac{E}{c^{2}}%
\right) \\
p_{y}^{\prime } &=&k_{y} \\
p_{z}^{\prime } &=&k_{z} \\
E^{\prime } &=&\frac{1}{\sqrt{1-(v/c)^{2}}}\left( E-vp_{x}\right) .
\end{eqnarray*}
applied to the relativistic theory of photons
\begin{equation*}
E=hf=pc=0/0\qquad p=h/\lambda
\end{equation*}
where $\lambda $ is the wave length, de Broglie obtained
the following relativistic equation for relativistic waves:
\begin{equation*}
\mathbf{p}=\hslash \mathbf{k}\qquad E=\hslash \omega .
\end{equation*}
This is how the wave-particle duality of quantum
mechanics was introduced. 

It was unfortunate that all of this was done
without noticing the following contradiction coming from the
relativistic theory of photons:
\begin{equation*}
E=\sqrt{(cp)^{2}+(m_{0})^{2}c^{4}}=cp=\frac{m_{0}vc}{\sqrt{1-\left( \frac{v}{%
c}\right) ^{2}}}=\frac{0}{0}cv=c^{2}hf=hf.
\end{equation*}

The reason why electromagnetic wave equations are invariant under
the Lorentz transformations is because Lorentz transformations are
transformations obtained within the context of Maxwell's electromagnetic
field theory.

Moreover, as we have shown above, Lorentz transformation fails to
map axioms of Newton's dynamics into axioms of Newton's dynamics. What does this
mean? A naturally expected answer would be that the concept of relativity is
not applicable to dynamics as it violates the law of action-reaction. But as we
have seen in the first section, Galilean transformation, which also is a
coordinate transformation associated with inertial reference frames,
conserves the second law and the gravitational law. The only deficiency of
the Galilean transformation we know of is that it violates the third law of dynamics.

This should mean that Lorentz transformations are more problematic
than Galilean transformations though both of them are invalid as they are
mathematical representations of the invalid concept of relatively moving
reference frames which violates the third law. 

However, in practice, in
most cases of dynamics around us, we can assume that one mass is way to
large to be able to ignore the reaction on it from a smaller mass, and so,
the conservation of the second law and gravitational law is enough to obtain a 
reasonable approximation. The action-reaction problem becomes important when we
consider the astronomical scale masses under gravitational acceleration.

\textbf{Remark (6)}
\textit{It is a tradition of pure mathematics and logic to
make sure that it is properly understood that a result must be always
accompanied with restrictions imposed on the derivation of it. Coordinate
transformations are applicable only to inertial frames. No accelerating
frames should be transformed. Yet in the discussion of transformation of
waves we are completely forgetting that physical waves come from
acceleration. This type of errors appear in many places in theoretical
physics, unfortunately. For example the well tested concept of centrifugal
force comes from the abusive use of a reference frame of an orbiting body.}

\section{Lorentz transformation and speed $c$}
What does it mean that Einstein proved that all axioms of
Maxwell's electromagnetic field theory are Lorentz invariant? This
transformation was built only for the electromagnetic field theory by Lorentz as a solution to the problem that the Michelson-Morley experiment introduced into the theory of electromagnetic
fields. That $c$ is the speed of light wave which is
electromagnetic wave in other reference frame gave an advantage. In other
theories of physics where the speed of light is not the issue, it is hard to
imagine that this $c$ will play the role it played in the
Maxwell's electromagnetic field theory. Therefore, not surprisingly,  Lorentz transformation fails to preserve the law of gravity and the
second law of classical dynamics. 

Maxwell's electromagnetic field theory is incomplete for a description of
electrodynamics. Electrodynamics assumes classical dynamics as an underlying
theory. This part was omitted by Maxwell and proponents. 

This problem surfaced explicitly in Lorentz force. Lorentz force, which depends also upon the speed,
contradicts the second law, which asserts that the force should be dependent
upon acceleration, not speed.

On the one hand the Galilean transformation preserves the second law
and the gravitational law, and on the other hand Lorentz transformation
fails to do so. This important fact has never been noticed before. 

Einstein faced problems in showing the invariance of all
axioms of Maxwell under the Lorentz transformation. Some of the axioms had
to be translated into equivalent ones. This is rather expected. The
problem is that $\mathbf{J}$ is not a force field. It is a
different monstrosity that was challenged when Maxwell put it in his
axiomatic theory of electromagnetic fields. $\mathbf{J}$ is
defined in terms of $\mathbf{v}$. Lorentz transformation has a
problem with transforming $\mathbf{v,}$ more specifically,
relativistic addition of speeds. The argument goes as follows:

Assume three inertial frames $F1$, $F2$ and $F3$. Let $v$ and $v\prime $ be the mutual speed
between $F1$ and $F2$, and $F2$ and $F3$, respectively. As these speeds are used to define the gamma factor, they must be pre-relativistic speeds, i.e.
classical speeds. So, $v+v^{\prime }$ is the mutual speed between $%
F1$ and $F3$.

The Lorentz transformation $L$ between $F1$ and $F3$ is defined as
\begin{equation*}
x^{\prime }=\frac{(x-(v+v^{\prime })t)}{\sqrt{1-(v+v^{\prime })^{2}/c^{2}}}%
,\quad y^{\prime }=y,\quad z^{\prime }=z,\quad t^{\prime }=\frac{\left( t-%
\frac{(v+v^{\prime })x}{c^{2}}\right) }{\sqrt{1-(v+v^{\prime })^{2}/c^{2}}}.
\end{equation*}
Let $L$ and $L^{\prime }$ be Lorentz
transformations from $F1$ to $F2$, and from $F2$ to $F3$ 
\begin{equation*}
x^{\prime }=\frac{(x-vt)}{\sqrt{1-v^{2}/c^{2}}},\quad y^{\prime }=y,\quad
z^{\prime }=z,\quad t^{\prime }=\frac{\left( t-\frac{vx}{c^{2}}\right) }{%
\sqrt{1-v^{2}/c^{2}}}.
\end{equation*}
\begin{equation*}
x^{"}=\frac{(x^{\prime }-v^{\prime }t^{\prime })}{\sqrt{1-(v^{\prime
})^{2}/c^{2}}},\quad y"=y^{\prime },\quad z"=z^{\prime },\quad t"=\frac{%
\left( t^{\prime }-\frac{v^{\prime }x^{\prime }}{c^{2}}\right) }{\sqrt{%
1-(v^{\prime })^{2}/c^{2}}}.
\end{equation*}
respectively. It is clear that $L \neq L^{\prime }\circ L$ where $L^{\prime }\circ L$ is the mathematical composition of $L$ and $L^{\prime }.$

It is contested by the mainstream that in the special theory of
relativity calculating $v+v^{\prime }$ is the wrong thing to do.
It is claimed that this addition should be replaced by the relativistic addition
(transformation) of speed 
\begin{equation*}
v\oplus v^{\prime }=\frac{v-v^{\prime }}{1-vv^{\prime }/c^{2}}.
\end{equation*}
This is to say that $L$ should be 
\begin{equation*}
x"=\frac{(x-(v\oplus v^{\prime })t)}{\sqrt{1-(v+v^{\prime })^{2}/c^{2}}}%
,\quad y"=y,\quad z"=z,\quad t"=\frac{\left( t-\frac{(v\oplus v^{\prime })x}{%
c^{2}}\right) }{\sqrt{1-(v+v^{\prime })^{2}/c^{2}}}.
\end{equation*}
If so, then why is $v+v\prime $ not replaced by $v\oplus
v^{\prime }$ in the gamma factor? Also we have a problem with $(v\oplus v^{\prime })$ in the Lorentz
transformation as Lorentz transformation is the agent that introduces 
the concept of relativity and one cannot use already relativistic concept $%
v\oplus v^{\prime }$ to define such a transformation. This is a
conceptual vicious circularity.
The shear reason that this kind of addition works for composing Lorentz transformations does not justify its use. Unfortunately this version of $L$ still fails $L=L^{\prime }\circ L.$ Mathematically, Lorentz
transformations are linear transformations from 4D space to itself. So, it
is highly irregular that algebraic composition of such transformations is
not the desired transformation.

Conceptually, $(v+v^{\prime })$ is the classically measured
speed. The reason why $(v\oplus v^{\prime })$ was introduced in the 
numerator part is because the classical addition $(v+v^{\prime })$
does not work for relativistic addition of speeds. As the relativistic
addition $(v\oplus v^{\prime })$ is introduced by a relativistic
argument, it is viciously circular to use this relativistic version in the
gamma factor. 

Moreover, mathematically we have a problem too. It is naturally
expected that $x"/t"$ will serve as the observed speed $v\oplus
v^{\prime }.$ There is no way to prove that they are equivalent.
Here is a simple calculation that leads to this conclusion. 
\begin{equation*}
\frac{x"}{t"}=\frac{(x-(v\oplus v^{\prime })t)}{\left( t-\frac{(v\oplus
v^{\prime })x}{c^{2}}\right) }=\frac{c^{2}(x-(v\oplus v^{\prime })t)}{%
c^{2}t-(v\oplus v^{\prime })x}=\frac{c^{2}x-c^{2}(v\oplus v^{\prime })t}{%
c^{2}t-(v\oplus v^{\prime })x.}
\end{equation*}
Note that even if we replace the classical $v+v\prime $ in 
gamma factor with relativistic $v\oplus v^{\prime }$, the above
calculation holds.

Now assume that 
\begin{equation*}
\frac{c^{2}x-c^{2}(v\oplus v^{\prime })t}{c^{2}t-(v\oplus v^{\prime })x.}%
=v\oplus v^{\prime }.
\end{equation*}
Then we have 
\begin{eqnarray*}
c^{2}x-c^{2}(v\oplus v^{\prime })t &=&(v\oplus v^{\prime })(c^{2}t-(v\oplus
v^{\prime })x. \\
c^{2}x &=&2c^{2}(v\oplus v^{\prime })t-(v\oplus v^{\prime })^{2}x \\
c^{2}x+(v\oplus v^{\prime })^{2}x &=&2c^{2}(v\oplus v^{\prime })t \\
\frac{x}{t} &=&\frac{(v\oplus v^{\prime })^{2}}{c^{2}+(v\oplus v^{\prime
})^{2}}.
\end{eqnarray*}
There is no reason for this to be true.

Empirically speaking, even though it could be possible to measure $v,$ what about $v\prime $? It is next to impossible to
measure $v+v\prime $, is it not? If $v$ is the speed of a
star A moving away from us in distance and if $v\prime $ is the
speed of another star B moving away from A, then how can we measure the $%
v\prime $ and so $v+v\prime $?

In case of axioms of electromagnetic fields, there is no $\mathbf{v}$ involved and this is why there was no difficulty for the invariance of the field equations of Maxwell. 

Despite all of this, there are some positive results concerning the
transformation of wave equations. Unlike Galilean transformation, Lorentz
transformation maps electromagnetic wave equations into electromagnetic wave equations. The reason for this is straightforward. This transformation assumes that the constant $c$ is
the speed of light in any inertial reference frame.

\textbf{Remark (7)}
\textit{Again, the above claim is false. As we will see later in 
\textbf{Section 6}, the claim that Lorentz transformations map wave
equations into wave equations is false even for electromagnetic wave
equations.}

\section{Is Schr\"{o}dinger's wave equation relativistic?}
\subsection{De Broglie relation}
De Broglie obtained the following relativistic transformation of a
plane wave for a wave invariant under the Lorentz transformation
(we call it a ``relativistic wave"):
\begin{equation*}
k_{x}^{\prime }=\frac{1}{\sqrt{1-(v/c)^{2}}}\left( k_{x}-v\frac{\omega }{%
c^{2}}\right) ,\text{\quad }k_{y}^{\prime }=k_{y},\text{\quad }k_{z}^{\prime
}=k_{z},\text{\quad }\omega ^{\prime }=\frac{(\omega -\nu k_{x})}{\sqrt{%
1-(v/c)^{2}}}
\end{equation*}
where $\mathbf{k}=(k_{x},k_{y},k_{z})$ is the wave vector
and $\omega $ is the frequency. We denote the wave number $
\left\vert \mathbf{k}\right\vert $ by $k$. So, $%
k=\left\vert \mathbf{k}\right\vert .$ This restriction to
relativistic waves is in place because otherwise the wave phase $\mathbf{k}\cdot 
\mathbf{r}-\omega t$ would not be invariant under the Lorentz
transformation. Using the analogy between this and the momentum-energy
transformation
\begin{equation*}
p_{x}^{\prime }=\frac{1}{\sqrt{1-(v/c)^{2}}}\left( p_{x}-v\frac{E}{c^{2}}%
\right) ,\text{\quad }p_{y}^{\prime }=p_{y},\text{\quad }p_{z}^{\prime
}=p_{z},\text{\quad }\omega ^{\prime }=\frac{(E-\nu p_{x})}{\sqrt{1-(v/c)^{2}%
}}
\end{equation*}
where $\mathbf{p}=(p_{x},p_{y},p_{z})$ is the momentum
vector and $E$ is the energy, de Broglie proposed the following
association between a particle and a wave (called matter wave): 
\begin{equation*}
\mathbf{p}=\hslash \mathbf{k}\quad E=\hslash \omega
\end{equation*}%
where $\hslash $ is a constant. It is called de Broglie
relation. Though it resembles Einstein's particle-wave duality 
\begin{equation*}
E=hf=pc\quad p=h/\lambda
\end{equation*}%
where $\lambda $ is the wave length, there is a fundamental
difference.

There are several issues to be discussed.
\begin{enumerate}
\item Unlike the photon-light duality where the speed of photon and
that of light are equal, the phase speed of matter wave and the speed of
particle can be different.

\item De Broglie further assumed that, associated with a particle
with speed $V$, was a wave having phase speed $w=\omega /k$. This association requires further explanation.

\item De Broglie also assumed that the energy in the wave traveled along
with a group speed $v_{g}=d\omega /dk$, which was identical to the
particle's speed $V$. Here it is not quite clear what did he mean 
by ``energy" in the wave. The de Broglie relation above is only a hypothesis based upon the above mentioned analogy of wave
vector-frequency transformation and momentum-energy transformation.
Certainly this does not yield the concept of energy in the wave.
\end{enumerate}

De Broglie assumed that $c^{2}(\omega /c^{2}-k^{2})$ is invariant under the relativistic transformation as an analogy to the
relativistic invariance of $c^{2}(E/c^{2}-p^{2})$. Thus, 
\begin{equation*}
c^{2}(\omega /c^{2}-k^{2})=constant=C.
\end{equation*}
From this, it follows that
\begin{equation*}
2\omega /c^{2}\frac{d\omega }{dk}-2k=0.
\end{equation*}
This leads to 
\begin{equation*}
v_{g}=\frac{d\omega }{dk}=\frac{c^{2}k}{\omega }.
\end{equation*}
As the phase speed is $w=\omega /k,$ we have 
\begin{equation*}
v_{g}=\frac{c^{2}}{w}.
\end{equation*}
It now follows that either the phase speed $w$ or the group
speed $v_{g}$ could exceed $c$, but not both. We do not
know what this means for the special theory of relativity, which asserts that
nothing moves faster than the speed $c$.

All of this is relative to the \textit{analogy-based hypothesis} that a particle with
speed $v$ has a wave dual called matter wave whose group speed is  
$v_{g}$ and whose phase speed is $w=\omega /k$. A
particle in motion carries energy and so it is expected that the wave dual
of this particle also carries energy of the same amount if the energy
conservation law is to be respected. But according to wave mechanics, for a
wave to carry energy it has to have wave medium. A concern we have is that
de Broglie's wave is a mathematical wave that appears to have no wave
medium, just like that electromagnetic wave carries energy without having
wave medium. We have already pointed out that electromagnetic field which
carries electromagnetic waves is a fiction, a counter-factual modality that plays no ontological role in physics. So, what happened to the energy issue
of the matter wave? This is not the case with de Broglie waves.

However, it is not clear why we have to choose Lorentz
transformed version over Galilean transformed version. That the
Galilean transformation of a wave function is a wave function, seems to
suggest that the theory of Lorentz transformation of wave functions is
rather self-serving. That Lorentz transformation came from time dilation 
and length contraction, which causes paradoxes (contradictions), seems to
suggest that there are more fundamental things that have to be reexamined in
the theory of Lorentz transformations. Indeed, almost all
waves that wave mechanics works on are not relativistic. The only familiar
waves that are relativistic are the electromagnetic waves. But this is overshadowed by the fact that the electromagnetic theory, which gave birth to
the electromagnetic waves, is not relativistic either. The most basic axioms of
this theory, the Coulomb's laws, are not relativistic as we have established
above. So, the claim that electromagnetic waves are relativistic strongly suggests
that the theory of electromagnetism is inconsistent.

\textbf{Remark (8)}
\textit{As we will see later in \textbf{Section
6}, the claim that Lorentz transformations map wave equations into wave
equations is false even for electromagnetic wave equations.}

\subsection{Schr\"{o}dinger's wave equation}
Schr\"{o}dinger used Hamilton's energy dynamics for the particle
theory and applied de Broglie's pilot wave theory to produce a wave-particle
duality that looks after the energy issue of de Broglie's relation.\cite{Sch}\cite{DeB}

All waves propagated along the $x$-axis obey the following
wave equation 
\begin{equation*}
\frac{\partial ^{2}\Psi }{\partial x^{2}}=\frac{1}{\omega ^{2}}\frac{%
\partial ^{2}\Psi }{\partial t^{2}}
\end{equation*}
where $\Psi (x,t)$ is the wave function and $\omega $ is the wave speed.

Here, we consider the wave function $\Psi $ whose square
yields the probability of locating a particle at any point in the space. We
consider only systems whose total energy $E$ is constant and whose
particles move along the $x-$axis and are bound in space. Then the
frequency associated via the de Broglie relation, which is hypothetical and relativistic, with the bound particle is also constant, and
we can take the wave function $\Psi (x,t)$ to be
\begin{equation*}
\Psi (x,t)=\psi (x)f(t).
\end{equation*}
As the frequency is assumed to be precisely defined, 
\begin{equation*}
f(t)=\cos 2\pi \nu t.
\end{equation*}
So, we have 
\begin{equation*}
\frac{\partial ^{2}\psi }{\partial x^{2}}=-\left( \frac{2\pi }{\lambda }%
\right) ^{2}\psi =-\left( \frac{p}{h}\right) ^{2}\psi
\end{equation*}
where the wave length is $\lambda =\omega /\nu $ and the
momentum of the particle is $p=h/\lambda .$

We take the particle of mass $m$ to be interacting with its 
surroundings through a potential-energy function $V(x)$. The total
energy of the system is given by 
\begin{equation*}
E=E_{k}+V=\frac{p^{2}}{2m}+V
\end{equation*}
where $E_{k}$ is the kinetic energy of the particle. Then
we have 
\begin{equation*}
p^{2}=2m(E-V)
\end{equation*}
and we have 
\begin{equation*}
\frac{\hslash ^{2}}{2m}\frac{\partial ^{2}\psi }{\partial x^{2}}+(E-V)\psi
=0.
\end{equation*}
So,
\begin{equation*}
-\frac{\hslash ^{2}}{2m}\frac{\partial ^{2}\psi }{\partial x^{2}}+V\psi
=E=i\hslash \frac{\partial \psi }{\partial t}.
\end{equation*}
This equation is called (non-relativistic) Schr\"{o}dinger wave
equation as the energy equation involves non-relativistic mass $m$ and it is not invariant under the Lorentz transformation. This does not mean
that quantum mechanics is a non-relativistic theory however. The derivation of Schr\"{o}dinger wave equation involved de Broglie relation which is nothing but
a relativistic theory.


We observe an issue with the above argument. It is claimed that
\begin{equation*}
\frac{\partial ^{2}\Psi }{\partial x^{2}}=\frac{1}{\omega ^{2}}\frac{%
\partial ^{2}\Psi }{\partial t^{2}}
\end{equation*}
is the wave equation and  
\begin{equation*}
-\frac{\hslash ^{2}}{2m}\frac{\partial ^{2}\psi }{\partial x^{2}}+V\psi
=E=i\hslash \frac{\partial \psi }{\partial t}.
\end{equation*}
is given as its example. This obviously becomes a wave equation only when $V=0$.

As pointed out above, Schr\"{o}dinger's wave equation is
in fact not a wave equation as it is irreconcilable with the classical
equation for waves. A further observation of its form indicates similarities
between Schr\"{o}dinger's equation and the diffusion equation used in
describing density fluctuations in materials due to diffusion. The diffusion
equation is given as
\begin{equation*}
\frac{\partial \phi (x,t)}{\partial t}=\nabla D(\phi ,x)\phi (x,t)
\end{equation*}
where $\phi (x,t)$ is the density of the diffusing
material at position $x$ and time $t$, and $D(\phi ,x)$ is
the diffusion coefficient for density $\phi $ at position $x$. When $D$ is constant, the equation reduces to
\begin{equation*}
\frac{\partial \phi (x,t)}{\partial t}=D\nabla ^{2}\phi(x,t).
\end{equation*}
which is a partial differential equation with first derivative in
time and second derivative in position, just like the Schr\"{o}dinger's
equation. This particular form of diffusion equation was proposed by Fourier
in 1822 to describe the heat distribution in a given region of a material
over a particular time and hence is sometimes referred to as
``heat equation".\cite{Fou}

A crucial difference between the Schr\"{o}dinger equation and the
diffusion equation is that the coefficient in the latter ($D$) is
real, while in the former it is complex. Consider for instance the equation
for a free particle:
\begin{equation*}
\frac{\partial \psi (x,t)}{\partial t}=\frac{i\hbar }{2m}\nabla ^{2}\psi
(x,t).
\end{equation*}
This difference makes the solutions to the diffusion equation decay with
time (gradient), while the solutions to the Schr\"{o}dinger's equation oscillate
(wave).  Note however that originally, before the Born interpretation became common, Schr\"{o}dinger attempted to interpret the wavefunction as electronic charge distribution in space (with charge density at position $x$ and time $t$ proportional to $|\psi|^2$): ``the charge of the electron is not concentrated in a point, but is spread out through the whole space [...] the charge is nevertheless restricted to a domain of, say, a few Angstroms, the wavefunction $\psi$ practically vanishing at greater distance from the nucleus."\cite{Sch} This would suggest his treatment of charge density as having a character of a radiation, indicating certain gradient properties.

It is unfortunate that the name ``wave
equation" became the common name for Schr\"{o}dinger's
equation, thereby confusing the classical wave equation with a formalism
lacking grounding in ontology. It appears that the Schr\"{o}dinger
equation is an attempt at merging the concept of wave-particle duality with
the treatment of electronic charges in terms of density distribution.
Regarding the former, Schr\"{o}dinger himself had reservations assuming the
veracity of matter waves, justifying the concept by stating that neglecting
de Broglie's waves leads to serious difficulties in atomic mechanics.
Regarding the latter, Schr\"{o}dinger himself noticed that this
interpretation of wavefunction does not work for systems of multiple
electrons.\cite{Sch2}

Nevertheless, Schr\"{o}dinger was aware of this problem and tried, unsuccessfully, to make
his wave equation relativistic. Later, Gordon, Klein and Dirac attempted to resolve this problem in the development of the quantum electrodynamics.

With all of this, it is clear why Schr\"{o}dinger failed to show
that his wave equation for particles is relativistic. To make the matter
even worse, Schr\"{o}dinger's equation is not relativistic, and thus a  quantization of such an equation is
impossible because de Broglie's quantization of waves worked only for
relativistic waves.

Instead of relativising Schr\"{o}dinger's
wave equation, Gordon and Klein quantized relativistic energy-momentum equation of
Einstein by replacing energy variable and momentum variable with
quantum energy operator and quantum momentum operator. This however does not make the Schr\"{o}dinger wave equation relativistic and therefore does not compensate for the deficiency stated above.

\textbf{Remark (9)}
\textit{The energy-momentum relation is a
consequence of the relativistic energy equation $e=mc^{2}$ which is
false. Here, $m$ is the relativistic mass $m_{0}/\sqrt{1-(v/c)^{2}}
$which is obtained through a thought experiment that assumed that $v$ is constant. With this Einstein obtained relativistic second
law $P=mv.$ By taking a time derivative, Einstein then obtained the
relativistic second law $F=dP/dt=vdm/dt+mdv/dt.$ This lead him to 
conclude that $e=mc^{2}.$ Unfortunately, $v$ is a 
constant, which leads to $e=0$ instead. As
Einstein pointed out, if $e=mc^{2}$ fails, the entire modern physics
fails. }

Dirac took advantage of the Gordon-Klein equation and derived
a relativistic theory of electrons which yielded the positron and opened a
gate to quantum electrodynamics which is considered the
most successful theory of physics in history.

\section{Are wave equations really relativistic?}
Now we have reached the point where the question has to be asked whether the wave equations really represent waves that appear in physics
correctly. Another question is whether the Lorentz transformation
which seemingly maps electromagnetic wave equations to electromagnetic
wave equations does so with ontological background.

The Galilean
transformation fails to conserve electromagnetic wave equations and the 
Lorentz transformation conserves electromagnetic wave equations. It seems to
be the only reason why Lorentz transformation replaced Galilean
transformation. The Galilean relativity theory was rejected (except the
faulty interpretation of the Michelson-Morley experiment) because it failed
to map electromagnetic wave equations to electromagnetic wave equations. So,
if can be safely said that as far as the wave theory is concerned, it was the
failure to conserve electromagnetic wave equations which dethroned the
Galilean transformation.

Here we have to ask whether the wave equations
are the basic axioms of physical theories. Clearly not. They are the product
of the basic axioms under certain circumstances. So, logically there is no
convincing reason why such secondary equations must be conserved under  coordinate transformations.

But to make the argument more articulate, let us discuss the issue
in a more general setting.
\begin{eqnarray*}
\frac{\partial \psi(x',t')}{\partial x} &=& 
\frac{\partial \psi(x',t')}{\partial x'}\frac{\partial x'}{\partial x} + \frac{\partial \psi(x',t')}{\partial t'}\frac{\partial t'}{\partial x} \\ &=&
\frac{\partial \psi(x',t')}{\partial x'}\frac{\partial \gamma(x-vt)}{\partial x} + \frac{\partial \psi(x',t')}{\partial t'}\frac{\partial \gamma(t-\frac{vx}{c^2})}{\partial x} \\ &=&
\gamma\frac{\partial \psi(x',t')}{\partial x'} - \frac{\gamma v}{c^2}\frac{\partial \psi(x',t')}{\partial t'}
\end{eqnarray*}
Similarly,
\begin{eqnarray*}
\frac{\partial \psi(x',t')}{\partial t} &=& 
-\gamma v\frac{\partial \psi(x',t')}{\partial x'} + \gamma \frac{\partial \psi(x',t')}{\partial t'}
\end{eqnarray*}
Then, 
\begin{eqnarray*}
\frac{\partial^2 \psi(x',t')}{\partial x^2} &=& 
\left(\gamma\frac{\partial}{\partial x'} - \frac{\gamma v}{c^2}\frac{\partial}{\partial t'} \right)\left(\gamma\frac{\partial}{\partial x'} - \frac{\gamma v}{c^2}\frac{\partial}{\partial t'} \right) \\ &=&
\gamma^2 \frac{\partial^2}{\partial x'^2} -2\frac{\gamma^2 v}{c^2}\frac{\partial^2}{\partial x' \partial t'} + \frac{\gamma^2 v^2}{c^4}\frac{\partial^2}{\partial t'^2}
\end{eqnarray*}
And similarly, 
\begin{eqnarray*}
\frac{\partial^2 \psi(x',t')}{\partial t^2} &=& 
\gamma^2 v^2 \frac{\partial^2}{\partial x'^2} -2\gamma^2 v \frac{\partial^2}{\partial x' \partial t'} + \gamma^2 \frac{\partial^2}{\partial t'^2}
\end{eqnarray*}
With this, the wave equation now becomes
\begin{equation*}
\gamma ^{2}\frac{\partial ^{2}}{\partial x^{\prime 2}}-2\frac{\gamma ^{2}v}{%
c^{2}}\frac{\partial ^{2}}{\partial x^{\prime }\partial t^{\prime }}+\frac{%
\gamma ^{2}v^{2}}{c^{4}}\frac{\partial ^{2}}{\partial t^{\prime 2}}=\frac{1}{%
\omega ^{2}}\left( \gamma ^{2} v^2\frac{\partial ^{2}}{\partial x^{\prime 2}}%
-2\gamma ^{2}v\frac{\partial ^{2}}{\partial x^{\prime }\partial t^{\prime }}%
+\gamma ^{2}\frac{\partial ^{2}}{\partial t^{\prime 2}}\right) .
\end{equation*}
This is valid only under the condition $v=c=\omega .$ The
second equality comes from the fact that $\omega $ is the wave speed. The
first equation implies that the frame speed is $c$ which is not
possible in the special theory of relativity. This means that Einstein's claim
that the electromagnetic wave equation is invariant under the Lorentz
transformation is invalid. It is a well understood fact that there is no
reference frame for light at the pain of contradiction.

We summarize the results thus far as follows:

\textbf{Conclusion (1)}
\textit{Lorentz transformation fails to conserve all wave equations
including electromagnetic wave equations.}

\textbf{Conclusion (2)}
\textit{Lorentz transformation serves no imaginable purpose. It fails
the conservation of the third law, that of the second law, that of
gravitational law, that of Coulomb's law. Now we also know that it does not conserve
even the electromagnetic wave equations. This removes the claim that Einsteinian relativity theory is more appropriate than Galilean
relativity theory. Naturally, Lorentz transformation does not conserve wave functions either.}

\textbf{Conclusion (3)} 
\textit{All of this is a consequence of the wrong
interpretation of Michelson-Morley's experiment. As we demonstrated previously \cite{Kan3},
the Michelson-Morley experiment showed that one cannot detect $v$ in $c+v$ in the way we measure the speed of light. Hence it appears that the problems in modern physics started from the Michelson-Morley
experiment.}

Galilean transformation conserves all of the basic laws
and constructs of physics except the third law and the wave equation. The
only issue with this transformation is that it is based upon the faulty
concept of moving reference frames. To see the problem with moving reference frames, assume a train runs on
a track. When the tip of the train's power pole touches the power line at
point A, a spark occurs at A. An observer located in the train straight down
from the tip A of the power pole will observe that this light comes straight
down to him/her from point A. But as point A also is a stationary
point of the power line, the observer will also see that the light reaches him/her
diagonally from point A on the power line.\cite{Kan2}

Mathematically this problem can be explained as follows: in geometry
one cannot move any point in geometric spaces as doing so breaks the metric
structure of geometric spaces. One cannot move a point 5 to the position
of point 3 and \textit{vice versa} as this breaks the metric topology of the real
number line. This is why Newton did not attempt to move any geometric points. Instead he
reduced a physical body to a point body and moved it inside a geometric
space. If we cannot move even a single point in a geometric space, how can
we move the entire space itself inside another space? If we move a point 5 on
a real line then what is the function that describes such motion?

Topologist Ren\'{e} Thom pointed out there is no point in
geometry. In geometry we must assume that mysterious linear ordering among
real points. This makes the geometry a continuum. Mathematical logician (the
founder of model theory) Abraham Robinson expressed this in terms of
infinitesimals. Points are all ``glued" together by invisible infinitesimals. In
the end, standard real analysis and infinitesimal calculus do the same
thing.

\section{Relativistic transformations and 4D spacetime}
A motion in the 3D space is a function $f(t)=(x,y,z)$. This
can be expressed as a line graph in the 4D spacetime. If the speed of the
motion is constant, the graph is a straight line, and if it is under
acceleration, then the graph is a curved line. When we apply a Galilean
transformation to this graph, then the resulting graph is a translated line.
\begin{equation*}
f(t)=(x-vt,y,z).
\end{equation*}

However, when we apply Lorentz transformation to this graph, due to
time dilation and length contraction combined, the resulting graph
becomes incomprehensible. So, the resulting graph is unusable for
the purpose of physics. In symbols, the resulting ``motion" becomes the graph
of
\begin{equation*}
f\left( \frac{1}{\sqrt{1-(v/c)^{2}}}\left( t-\frac{vx}{c^{2}}\right) \right)
=\left( \frac{1}{\sqrt{1-(v/c)^{2}}}\left( x-vt\right) ,y,z\right) .
\end{equation*}
This is expected as under the Lorentz transformation time
and space coordinates are interdependent.

\section{Dirac's aether theory}
As can be seen in the vortex theory, which we will discuss in the
next section, the whole philosophy of aether theory is to ``squeeze out"
particles from a continuum. At the most fundamental level, as Ren\'{e} Thom pointed out, this is impossible as it destroys the
topology of the continuum. The difficulty the classical aether theory had
is naturally expected because of the nature of continuum. 

Dirac was the first who managed to create this paradigm upon the
quantum field which is the quantization of classical field using the
mathematical tool of Fourier expansion.\cite{Dir2} In this method, Dirac did not
create a geometric point as the quanta. He created a finite approximation of
infinite Fourier expansion as a particle. So, Dirac's particles are infinitary
objects described by waves. 

This project started with a new theory of photons proposed by Planck, which Planck himself did not take seriously and
presented as a purely mathematical convention arising from graph fitting as the
last resort to resolve the mystery of the blackbody radiation. Planck presented
the argument that if one accepts that the minimum energy carried by the
electromagnetic wave is $hf$, where $f$ is the frequency
of the wave and $h$ is a constant, which is now called the Planck
constant, the infamous blackbody radiation problem is resolved. So, Planck
proposed that the light wave of frequency $f$ carries waves as $nhf$, where $n$ is a natural number.

Under the assumption that the speed $v$ of light in
vacuum without conducting current is constant $c$, which came
from Maxwell's theory of electromagnetism, Einstein concluded that the mass
of Planck's particle (photon) must be $0$ to avoid the
relativistic energy
\begin{equation*}
e=mc^{2}=\frac{m_{0}}{\sqrt{1-(v/c)^{2}}}c^{2}
\end{equation*}
of the photon become undefined (or diverge) where $m$ is the
rest mass of the photon. With this convention, Einstein took the energy
equation for the photon to become
\begin{equation*}
e=0/0=hf.
\end{equation*}

\textbf{Remark (10)}
\textit{Einstein thought that since $0/0$ is equivalent to $0x=0$, and for the latter $x$ can be any number, $0/0$ can be
any number and he chose it to be $hf.$ However the difference between the two is such that the former involves
division by $0$, which is not allowed in mathematics, and the latter does not
involve it.}

As discussed above, this leads
to yet another contradiction. The relativistic energy equation $e=mc^{2}$ 
leads to the famous relativistic energy-momentum relation $e=\sqrt{(cp)^{2}+(m_{0}c^{2})^{2}}$ which in turn leads to the following
contradiction
\begin{equation*}
e=\sqrt{(cp)^{2}+(m_{0})^{2}c^{4}}=cp=\frac{m_{0}vc}{\sqrt{1-\left( \frac{v}{%
c}\right) ^{2}}}=\frac{0}{0}c^{2}=c^{2}hf=hf.
\end{equation*}

Logically speaking the real problem with the Planck-Einstein photon
theory is that the issue of blackbody radiation was an empirical refutation of the classical electromagnetic field theory of Maxwell. The convention Planck and Einstein presented, which turned out to be invalid  as we have shown, did not repair the deficiency of Maxwell's theory. No change was made to Maxwell's theory after the Planck's proposal. So, these two mutually contradicting theories were combined together to produce another theory that makes opportunistic choice. Namely, when it comes to most of the classical part of
electromagnetism, it uses the original Maxwell's theory and when it
comes to the issue of the light waves, it chooses the Planck-Einstein addition,
which contradicts Maxwell. 

Dirac does not appear to have known of this
fatal error of Planck-Einstein. But he was rightly unhappy with the \textit{ad hoc} nature of the process of
obtaining the equation $e=0/0=hf.$ He concluded that obtaining 
photons from electromagnetic wave equation is the wrong thing to do.
So instead, Dirac presented the photons through Fourier expansion of the
vector potential. In this way, he managed to obtain a richer theory of photons.
However, deducing photons through Fourier expansion of vector potential
lacks in ontology. Also, the quantization of the charges and currents in the Maxwell's theory remained to be reviewed. 

Dirac's solution to the problem of correctly quantizing charges
and particles was to rely upon the Schr\"{o}dinger wave equations. 
He found it impossible to quantize charges and particles as they are
already particles. According to the basic principle of wave-particle
duality, namely de Broglie relation, quantum particles must come from
waves. So, Dirac first converted particles such as charges into the Schr\"{o}dinger wave equations. Instead of going through von Neumann's quantization,
Dirac used Fourier expansion of the solutions of the wave equations to
create quantized particles. Particle interactions were modeled through
the interference of the wave equations derived from the particles. Through
this process Dirac obtained more particle varieties and more interesting
operators on particles such as the annihilation operators.

Unfortunately, as we have discussed in the section ``Is Schr\"{o}dinger's wave equation realtivistic?", Schr\"{o}dinger's wave equations
are not relativistic, meaning that they are not invariant under the Lorentz
transformation in general. This means that the claim of Dirac that his new
theory of quantum electrodynamics is relativistic is false as his
quantization uses Schr\"{o}dinger's wave equations. To make the matter even
more confusing, Schr\"{o}dinger's wave equation was obtained by  applying de
Broglie relation to classical Hamiltonian theory of particles and this
relation is relativistic upon the assumption that the wave of de Broglie is
relativistic (meaning Lorentz-invariant). Schr\"{o}dinger's wave
equation is not relativistic because Schr\"{o}dinger misunderstood what
de Broglie did. Schr\"{o}dinger used de Broglie relation to convert
classical Hamiltonian energy equation into a wave function. De Broglie did
not associate a particle to a wave equation however. Indeed, what he did was the 
opposite. He associated a particle having relativistic energy and relativistic
momentum with waves. His wave-particle duality is a one way association.
Moreover de Broglie had to assume that the wave equation in his theory has
to be relativistic, meaning that it is invariant under the Lorentz transformation.
For such a relativistic wave equation, he associated a particle with energy
and momentum.

In this way, the wave-particle duality of Schr\"{o}dinger's is fundamentally flawed, putting the invalidity of relativity theory
aside.

After all, as the relativity theory is inconsistent, there is no
point in considering whether Dirac's theory is relativistic or not.

Moreover, Dirac quantized electromagnetic fields, which are not
physical reality but a modality, through Fourier expansion to obtain photons.
This makes Dirac's photons suffer from the same conceptual obscurity as
Planck-Einstein's photons, which are also the product of quantizing (in a
different way) electromagnetic waves, which are modal waves.

Regarding Feynman's quantum electrodynamics, despite some
improvements such as leaving Hamiltonians behind and moving into the
Lagrangian, this theory did not resolve the problem associated with Schr\"{o}dinger's wave equations discussed just above. 

Gordon-Klein's quantization of invalid relativistic
energy-momentum equation of Einstein does not offer any solution to this
fundamental problem that Schr\"{o}dinger's wave equation is not
relativistic. Replacing relativistic energy variable and relativistic
momentum variable with energy operator and momentum operator in the faulty
relativistic energy-momentum relation is not what a 
quantization should entail. 

Below we posit some major questions regarding how quantum physics led to quantum electrodynamics.

\begin{enumerate}
\item There are many concepts of quantization. Namely,
Planck-Einstein quantization of electromagnetic waves, de Broglie's
quantization of associating relativistic waves with a particle with
momentum and energy through analogy between the transformation of
relativistic waves and transformation of energy-momentum, Schr\"{o}dinger's quantization of converting classical particle equations into wave
equations using de Broglie's relation, Dirac's quantization of
electromagnetic fields through Fourier expansion, Dirac's quantization of
particles expressed as Schr\"{o}dinger wave equations through Fourier
expansion, Gordon-Klein quantization of Einstein's energy-momentum
relation, etc. Yet there is no study of how they are related. 

\item De Broglie's quantization, which
plays key role in many quantizations as listed above, is not properly
understood. This quantization works only for relativistic waves that are
invariant under the Lorentz transformation. As the momentum-energy of de
Broglie particle is related to the transformation of relativistic waves only
through analogy, we cannot find a way to obtain a wave that is
relativistic and represents the original particle. As Schr\"{o}dinger's wave
was created using this obscure de Broglie relation, the validity of it is
questionable. This makes Dirac's quantization of Schr\"{o}dinger's
wave questionable as well. 

\item When it comes to the Gordon-Klein equation, which is accepted as an
alternative to the failed attempt of making Schr\"{o}dinger's wave mechanics
relativistic, this is an attempt to quantize a relativistic relation in an
unprecedented way. Does replacing classical variables with corresponding
Hermitian operators make the classical theory quantum? 

\item It is important to ask these questions instead of experimentally try to verify the predictions of these incoherent theories
where the core discussion is based only upon analogy and the wrong assumption
that Schr\"{o}dinger's wave equations are relativistic. On the top of it, as
quantum theory is inherently probabilistic, its experimental
verification is highly compromised. The verification is done as the statistical
calculation of standard deviation. So, the claimed accuracy of the 
expensive experiments on particles is verified in the same way we
evaluate the accuracy of the prediction of the life span of automobiles.
\end{enumerate}

Regarding (iii) above: The Gordon-Klein equation does not conserve
probability, the conservation of which is a major requirement imposed by the usual
interpretation of quantum mechanics. Quantum mechanics interprets the square
of the modulus of a wave function's amplitudes as probabilities. For that
reason, Schr\"{o}dinger's equation was made to make sure that the
coefficients of wave functions were normalized at every point in time. This
unfortunately is not the case for the Klein-Gordon equation. It cannot therefore be seen as a valid replacement for a relativistic
version of Schr\"{o}dinger's equation. In order to conserve probability, a
time evolution equation needs to satisfy the following condition with
regards to a wave function
\begin{equation*}
\int \left\vert \psi (x,t)^{2}\right\vert dx=1
\end{equation*}
Furthermore, as the conservation must hold at any point in time, it
has to be independent of time evolution. This is to say that the Gordon-Klein
equation must satisfy the following equation as well
\begin{equation*}
\frac{\partial }{\partial t}\int \left\vert \psi (x,t)^{2}\right\vert dx=0.
\end{equation*}
Now consider the Gordon-Klein equation
\begin{equation*}
\frac{1}{c^{2}}\frac{\hslash ^{2}\partial ^{2}}{\partial t^{2}}\psi
(x,t)=(\hslash ^{2}\nabla ^{2}-m^{2}c^{2})\psi (x,t).
\end{equation*}
Since it involves the second derivative with regards to time, it is
clear that the first derivative term in the probability conservation
expression will in general not disappear. Hence, the expression will not
produce the required value $0$ and therefore the Gordon-Klein
equation clearly does not describe the probability wave that the Schr\"{o}dinger equation describes.

The most important issue is that relativity theory as per
Einstein is false and there is no point in trying to make classical
theories relativistic. Classical theories such as the theory of electromagnetism have their own problems. Relativity theory is a wrong answer to the problem
of classical electromagnetic theory. Considering the fact that the theory of relativity came from the wrong interpretation of the Michelson-Morley
experiment, the entire quantum theory must be reevaluated.

\section{Aether theory}

\subsection{Gravitational aether}
\textrm{Classical aether theory proposed by Descartes is yet another
example of a continuum medium producing atoms (particles) through type
lowering. Here a vortex, which is a substructure of the universal medium
aether, is supposed to be the atom which will induce the inter-atomic forces. We do not know how far we can push this idea forward as from the start this
idea leads to a contradiction. Here is a brief discussion on the
basic idea of Descartes on aether theory:}

\begin{enumerate}
\item \emph{Proposition}\textrm{: } \emph{``No empty space can exist, therefore space must be filled with matter."}\textrm{\ \quad \smallskip
Descartes is saying that there is no such thing as a geometric space like e.g. the 3D space then. As Newton made it clear, no matter what we place in a
geometric space, the space itself is a geometric space. Otherwise we cannot
even define a motion which is, as Newton said, a function from time to space. 
The other way of nailing down the problem is that by saying ``space must be
filled with matter" Descartes is already assuming that space is a container that
can be empty. Yet he is claiming that such thing does not exist. }

\item \emph{Proposition}\textrm{: } \emph{``Each part of this matter is
inclined to move in straight paths, but because parts are close to each
other, their interaction makes every part make circular motion. Each part
making this circular motion is called vortices. They are often called
`atoms'. Descartes also assumes that rough matter resists the circular
movement more strongly than fine matter."} \quad \textrm{It appears that
just this claim requires a massive amount of physics.} \textrm{This requires a fully
developed and articulate theory of fluid. It is clear that the theory
of fluid should be something much more advanced than particle-based
dynamics. In fact, we have a serious problem with the transition from
particle dynamics to fluid dynamics. Indeed, it is becoming increasingly clearer that fluid dynamics is a very different theory from particle dynamics.
So, it appears that before we venture into aether theory we must have a
solid understanding of what fluid mechanics is about. We certainly do not have a clear notion of fluid dynamics and its relation with particle dynamics yet. }

\item \emph{Proposition}\textrm{: } \emph{``Due to centrifugal force, matter
tends towards the outer edges of the vortex, which causes a condensation of
this matter there. The rough matter cannot follow this movement due to its
greater inertia---so, due to the pressure of the condensed outer matter, those
parts will be pushed into the center of the vortex. This inward pressure is
gravity." \ \quad }\textrm{There is no such thing as centrifugal force. This
is why this force is called a fictitious force. This is a good example of
how the violation of the principle of relativity of Galileo occurs when we
consider reference frames under acceleration. This is why
we do not allow reference frames under acceleration. The effect of the so-called centrifugal force appears only when we consider an object floating
inside a container that is rotating about a centre of the rotation. This
body tries to stay where it is when the centripetal force pulls the
container down. If a body is fixed to the body of the orbiting container, it
will not feel any centripetal force. After all this kind of situation is not
theorizable as the classical particle dynamics does not allow us to consider
things like orbiting container that has an inside space. To be precise, every
object must be a point object in classical dynamics. }
\end{enumerate}

\textrm{Upon the ideas of Descartes, Huygens presented a more articulate
vortex theory. The following is a short discussion on his work: }

\begin{enumerate}
\item \emph{Proposition}\textrm{: } ``\emph{Huygens assumed that the free moving
aether particles are pushed back at the outer borders of the vortex and
causing a greater concentration of fine matter at the outer borders of the
vortices. This causes the fine matter press the rough matter into the
center of the vortex." \quad\ }\textrm{It is not clear how this distribution
of the fine matter (aether particles) will occur. This argument requires a
full theory of particle-based fluid mechanics. It requires a very advanced
theory to explicate this process. More fundamentally, due to the very
concept of fluid, particle-based fluid is untenable. Particles and fluid cannot be unified. The former is discrete and the latter is continuous. As the
space is a continuum, no matter how densely we pack the particles, we still have
empty spaces in between the packed particles. The only aether we can think of must be continuum fluid. }

\item \emph{Proposition}\textrm{: } \emph{``According to Huygens the
centrifugal force is equal to the force that acts in the opposite
direction of the centripetal force."} \quad \quad \textrm{Again, this claim
needs a fully developed fluid theory. Huygens' definitions of centripetal force
and centrifugal force are different from standard Newtonian version.
Newton's version is simple and clear. There is no such thing as a centrifugal
force. It is a misunderstanding of the fact that a free body inside a
container will remain where it is despite the motion of the container under acceleration. So, there is no such thing as a centrifugal force. This ``force" arose when Newton's successors misunderstood Newton's theory and included the reference frame.}

\item \emph{Proposition}\textrm{: } \emph{``Huygens also assumed that ``bodies",
whatever they may be, must consist mostly of ``empty space" so that the
aether can penetrate them." \ \quad }\textrm{Huygens assumed that there is
no such thing as empty space.}\emph{\ }\textrm{Moreover, there is no
definition of what a body is.}

\item \emph{Proposition}\textrm{: } \emph{``He further concluded that the
aether moves much faster than the falling bodies." \ \quad \smallskip }%
\textrm{A more fundamental question is: what is causing the motion of aether
(aether particles) such as fine matters and rough matters?\ The same
question can be asked about the issue of the motion of bodies. }

\item \emph{Proposition}\textrm{: } \emph{``His theory could not explicate
Newton's law of gravity, the inverse square law. Huygens tried to deal with
this problem by assuming that the speed of the aether is smaller in greater
distance." \ \quad }\textrm{Again, the same problem as above. We do not know
what the speed of aether is until we learn what causes the motion of aether. }
\end{enumerate}

\textrm{The overall judgment on Huygen's aether theory is that it failed to
explicate the dynamics of aether itself. It appears to be something
even more complex than what we know as fluid dynamics. Fluid
dynamics is a derivative of Newton's dynamics that came with great compromises. Fluid
dynamics is continuum dynamics as fluid is a continuum. The compromises made include, for example, that ``pressure" is a
highly questionable derivative of Newton's force as a vector. In dynamics,
force is applicable only to a point matter because force is a pointed arrow.
This concept was extended from a point to an area or to a
volume, going backward of the direction Newton took to make physics possible,
which is to reduce a continuum body to a point body. This
compromise and Newton's mechanics combined created fluid dynamics. Therefore it is difficult to imagine that a theory of aether can be framed without using Newtonian
mechanics as it was the case with fluid dynamics.  }

\subsection{Electromagnetic aether}
\textrm{We have discussed the difficulty in producing continuum dynamics
(fluid dynamics) from particle dynamics of Newton. Aether theory starts with a super fluid structure called
aether and then, from the aether, induces particles called atoms.} \textrm{This
is yet another example of type lowering taking place in theoretical physics
as the fluid is treated as a continuum made from points. So, the trouble associated
with the type lowering manifests itself in any aether theory. }

\textrm{\ Maxwell was compelled to reject his aether theory and accept the
field theory for the theory of electromagnetism of Heaviside and Hertz as a shortcut to resolving 
the problem of nonlinearity. The problem here is that the theory of electromagnetic
fields is not ontological as the concept of a force field is
not reality. It is counterfactual modality, as we discussed above.
Moreover, it was not understood that the force field theory violates the 
law of action-reaction and that it is a modal theory. Newton was aware of this and did not use the concept of a force field.}

\textrm{ As we showed, Lorentz transformation does not conserve electromagnetic wave
equations and so by proof by contradiction we can conclude that it does
not conserve all of the Maxwell's axioms, contrary to the claim by Einstein
that all axioms of Maxwell are Lorentz-invariant. If all axioms of Maxwell
were relativistic as Einstein attempted to prove, then the electromagnetic wave equations must also  be 
relativistic. This is to say that \emph{as Maxwell's electromagnetic
wave equations are not relativistic, the theory of electromagnetic field as per Maxwell
is not relativistic.} This is consistent with Einstein's failure to prove that
Lorentz transformations conserve all axioms of Maxwell. }


\textrm{Now let us go back to the aether theory which Maxwell tried to build
in order to interpret axiomatic electromagnetic field theory. Maxwell was
reluctant to accept Heaviside's and Hertz's axiomatic approach of compiling
experimental lab results as the vector equations of electromagnetic fields.
From our view point, Maxwell was correct in this reluctance as we understand
that force fields violate the third law of Newton. Record
shows that even after accepting the axiomatic force field approach
in producing his axioms of electromagnetic filed theory, Maxwell still was
attempting to push forward with his aether theory. One of the reasons for this was that
his theory of electromagnetic waves required a medium as all waves of physics
need a medium, while the axiomatic theory does not provide it. }

\textrm{There are some factors that made it very difficult for this project of Maxwell's to
succeed. In order to succeed, we have to consider at least the following issues: }

\begin{enumerate}
\item \textrm{ Electromagnetic force must respect the law of action-reaction.
This is in conflict with the electromagnetic force fields which violate the law of 
action-reaction. Maxwell's field equations did not produce a solution here. This means that the right ontological electromagnetism theory, if any, shall
not agree with the description of Maxwell's electromagnetic field axioms,
which ignores the third law. }

\item \textrm{Exactly the same argument applies regarding modality. As
discussed above, force fields are not ontological reality. They are all
counterfactual modalities. However, the desired ontological theory of electromagnetism
cannot be a modality of any kind. }

\item \textrm{All of the above is to question whether the desired aether theory, if any, would be 
a modal theory or not. The answer naturally is ``no". A modal theory does not
describe physical reality. The concept of force fields must be rejected from
this point of view. There is no such thing as an 
electromagnetic wave as there is no such an ontological reality as a modal wave.
The correct view of what we call ``electromagnetic wave" is the transmission
at a distance of the vibration of electromagnetic force to a location where
there is a charge. There is no transmission in between. This is to say that in reality 
there is no such thing as electromagnetic waves. }
\end{enumerate}

\textrm{All of this suggests that trying to find an aether model for
Maxwell's electromagnetic field equations is futile. Instead, we should
focus on developing the unjustly abandoned Gauss-Weber's action at a
distance theory of electromagnetism \cite{Wes} which does not use problematic
field equations. The reason why Newton's dynamics is a little less
problematic is because it is not a force field-based theory.}

\textbf{Remark (11)}
\textit{Electromagnetic force depends on the
speed of charge in either the Maxwell-Lorentz formalism (known as the Lorentz
force) or the Gauss-Weber formalism. This makes electrodynamics
inconsistent as it violates the second law, which is a most important axiom of any dynamics.
Therefore it is not quite clear how we can put gravitational aether and
electromagnetic aether together. }

\textrm{From logical perspective, it is clear that an aether theory, if any,
would be more complex than the theory that the given aether theory attempts to
explicate. So, attempts to lay out an aether theoretic foundation of a physical
theory will tend to be viciously circular. And even if not, there is no
obvious way to verify such a meta-theory theoretically and empirically. One
may say that we can do that through an empirical verification of the theory
which the given aether theory is to lay foundation for. Then it is nothing but vicious
circularity. }

\section{Type lowering in mathematics}
The problem of type lowering which we discussed in the preceding in the context of theoretical physics also appeared in mathematics in more
acute forms. We will discuss some of them here.

\subsection{Scott's model of lambda calculus \cite{Sco}}
Church developed a symbolic reductional calculus called $\lambda $
\emph{-calculus} that described the theory of applying a function
of one variable to another such function. By defining natural numbers as a
special collection of such symbolic functions, Church simulated universal Turing
machine showing that his calculus has the same computational power as that
of Turing machines. But since its invention, this formal calculus needed a proof
that this reductional calculus is consistent. Dana Scott presented an
interpretation of this symbolic calculus by considering a set equation 
\begin{equation*}
D=[D\rightarrow D]
\end{equation*}
where the right hand side represents the set of all functions from
the set $D$ to itself. It is a well know fact that for
any set $D$, $[D\rightarrow D]$ is larger than $D$. So, Scott introduced a complete lattice structure and restricted
the elements of $[D\rightarrow D]$ to order continuous functions.
In this way he manage to cut down the size of $[D\rightarrow D]$
and establish a complete order isomorphism between the left hand side and the right hand 
side of the above equation, presenting the ``first model of $\lambda $-calculus". This success came with a price. Now we identify a
natural number with an element of $D$ which is infinitary. Therefore
in Scott's calculus we cannot decide if two natural numbers are equal. In Scott's model, if a term is reducible to another term, semantically, there are
many terms that do not syntactically represent natural
numbers but we cannot find that out using the syntactic reduction of the
calculus. Logicians call this kind of natural numbers recursive natural
numbers.

\subsection{Universal set theory CFG}
Consider the solution of the following set equation
\begin{equation*}
S=[S\rightarrow T]
\end{equation*}
where $T$ is the truth value set $\{true,false\}$. Unfortunately this equation has no solution as the right hand side
again is larger than the left hand side. Russel presented this problem as
the famous set as one (left side of the equation) and set as many (right
hand side of the equation) paradox. This tells us that the claim that a set can be
fully described by its characteristic function is not correct. This is yet
another paradox of set theory. The method Scott developed to solve $D=[D\rightarrow D]$ gives us a solution as the collection of order
continuous functions. But it is a well known fact that sets as
characteristic functions in mathematics are not ``order continuous" though
they are ``order monotonic". So, we have to solve the equation as the
collection of order monotonic functions from $S$ to $T$.
Apostoli and Kanda \cite{Apo} found a solution as a set of monotonic functions from $S$ to $T$. In this way the authors obtained the
first consistent universal set theory which has the set universe $S$. But this set theory, called CFG, has a drawback. CFG cannot identify two
sets on the basis of the ``extensional identity" which says that two sets are
equal if and only if they are made of exactly the same member sets. The so-called axiom of extensionality fails. It is replaced by the axiom
of indiscernibility which says that two sets are equal if and only if they
belong to exactly the same members of $S$. The loss of extensional
membership relation makes it unusable in the mathematics for working
mathematicians. This is yet another price we pay for type lowering.

\subsection{Type lowering in recursion theory \cite{Rog}}
Recursion theory is a branch of mathematical logic developed by G"{o}del in which we define computable partial functions of natural numbers
as functional programs over natural numbers. Using Turing machines, G\"{o}del showed how to calculate a natural number which uniquely represents a
functional program as a natural number. This process is called G\"{o}del
numbering of partial recursive functions. This process certainly is a type
lowering from the type of computable functions to natural numbers. Here each
computable function will be represented by infinitely many natural numbers
each of which represents a functional program that computes the computable
function. This implies that there are infinitely many recursive programs for
each computable function. However, at the pain of contradiction, given two
natural numbers, we can not computationally decide if the functions by these
two numbers are the same or not. So, we loose the identity of computable
functions.

\section{Type lifting in mathematics and particle-based physics}
Understanding all of these fundamental difficulties that the top down
approach creates, mathematicians took the bottom up approach as a better
methodology for building mathematical theories. A good example is the
development of the theory of real numbers. It goes as follows:

\begin{enumerate}
\item Natural numbers: closed under the operations $+$ and $\times .$

\item Integers: closed under the operations $+,\times $ and $-.$

\item Rational numbers are precisely the fractions $n/m$ of integers where $m\neq 0:$ closed under $+,-,\times $ and $\div .$ They are precisely the repeating infinite decimals.

\item Irrational numbers are non repeating infinite decimals. 

\item Real numbers are precisely the collection of all rational
numbers and irrational numbers. Real numbers are also closed under $+,-,\times ,$ $\div .$ Moreover, they are closed under bounded limit. 
\end{enumerate}

From this definition of real numbers we can prove that the real
numbers are a ``bounded complete ordered field." This is because the algebra $(
\mathcal{R},+,-,\times ,\div ,\leq )$ is an ordered
field with the linear ordering $\leq $ and is closed under bounded
limit. Here $R$ is the set of all real numbers. The
mentioned closure under operations properties can readily be proved except
the bounded limit which requires a little bit of work. 

This is a simplest way of developing the theory of real numbers so
that we can make it into a calculus (mathematical analysis). One cannot rely upon intuitive, simplistic
understanding of real numbers to develop calculus by replacing the concept
of limit by a geometric intuition. 

This type lifting (or bottom up) approach is based upon the same
philosophy of atomism in physics where the most basic physical entities are
atoms and from atoms we build more complex physical entities.

\end{document}